\documentclass[]{interact}

\usepackage{epstopdf}
\usepackage[caption=false]{subfig}

\usepackage{siunitx}
\usepackage{braket}
\usepackage{url}
\usepackage{xcolor}

\usepackage[numbers,sort&compress,merge]{natbib}
\bibpunct[, ]{[}{]}{,}{n}{,}{,}

\theoremstyle{plain}

\theoremstyle{definition}

\theoremstyle{remark}

\newcommand{\figref}[1]{Fig.~\ref{#1}}
\newcommand{\Figref}[1]{Figure~\ref{#1}} 

\newcommand{\wn}[0]{cm$^{-1}$}

\begin{document}


\title{State-selected preparation of molecular ions for precision measurements in radio-frequency traps}

\author{
\name{Daniel Y. Knapp and Maximilian Beyer \thanks{CONTACT M. Beyer. Email: m.beyer@vu.nl} }
\affil{Department of Physics and Astronomy, Vrije Universiteit Amsterdam, Amsterdam, The Netherlands}
}

\maketitle

\begin{abstract}
The application of mass-analyzed threshold ionization (MATI) for the state-selective preparation of molecular ions is presented. Based on photoexcitation of long-lived high-$n$ Rydberg states, molecular ions are prepared in a single rovibronic level by pulsed-field ionization. We present a theoretical analysis and a recipe for obtaining an optimal energy ratio between such selected ions and molecular ions in unwanted rovibronic states, created by direct photoionization. It is shown that the second-order chromatic aberration of a dc quadrupole bender can be used to isolate the state-selectively prepared molecular ions. The phase-space properties of ions prepared by MATI are ideally suited for axial injection into a linear radio-frequency trap. A modified approach for carrying out MATI within such an ion trap is also described.  

\end{abstract}


\section{Introduction}
Precision spectroscopy of molecular ions has emerged as a tool for testing fundamental physics and for developing molecular ion clocks. Applications include determining fundamental constants \cite{alighanbari2020a,patra2020a,schenkel2024a,alighanbari2025a,karr2016a,karr2023a,karr2025a}, searching for their possible variation \cite{schiller2005a}, and probing fundamental symmetries \cite{roussy2023a}.

Preparing ions in a single, well-defined quantum state is essential, both to raise signal-to-noise ratios in experiments with large ion clouds and to enable measurements with a single ion. Because molecules possess rotational and vibrational degrees of freedom, the population spreads over many levels, and so state-selective preparation methods are essential compared to atoms.

Electron bombardment is a common route to ion production, but offers little state selectivity \cite{weijun1993a}. It is often applied to thermal molecular samples, where the population is already distributed over many rovibrational levels. For heteronuclear diatomic and polyatomic ions, a non-zero electric dipole moment permits fluorescence on experimental timescales, and as a result, electronically or vibrationally excited populations tend to relax into the vibronic ground state. Nevertheless, black-body radiation at typical trap temperatures redistributes the population among rotational levels, even for light ions with large rotational constants \cite{biesheuvel2016a}.

Resonance-enhanced multiphoton ionization (REMPI) uses a resonant transition to excite a chosen rovibrational level of the neutral precursor. Because ionization accesses the continuum, selectivity in the ionic states is governed primarily by energy conservation. With a suitable resonance and an appropriate REMPI scheme, the lowest rovibrational levels of the ion can be loaded \cite{schmidt2020a, zhang2023a}. Compared to electron bombardment, the higher selectivity of REMPI arises from the much narrower energy spread of the laser compared to the electron beam. In addition, propensity rules associated with the autoionizing Rydberg states accessed in REMPI can bias the resulting distribution over ionic rovibrational levels \cite{ohalloran1988a}.

Without a state-selective source of ions, quantum-logic spectroscopy (QLS) offers a practical alternative \cite{wolf2016a,chou2017a,sinhal2020a,najafian2020b,chou2020a}. A single trapped molecular ion can be initialized in a chosen rovibrational state, repeatedly interrogated, and reused, thereby reducing the load on ion production by avoiding destructive detection. Starting from an unselective source, logic operations with a co-trapped atomic ion can be used to identify the target state and then isolate it for further study \cite{holzapfel2025a}.

An alternative strategy is to develop ion-production methods that are both highly selective and capable of high repetition rates. Merkt et al. pointed out in 1993 that the spectroscopic scheme underlying zero-kinetic energy photoelectron spectroscopy (ZEKE PES) \cite{muller-dethlefs1998a} and mass-analyzed threshold ionization (MATI) \cite{zhu1991a}, delayed pulsed-field ionization (PFI) of high-n Rydberg states, can be used to prepare ions in selected rovibrational levels \cite{merkt1993c}. In their work, hydrogen molecular ions were selectively prepared in different rotational levels of the electronic ground state with $v^+=2$. An extension to polyatomic molecules with a focus on infrared spectroscopy was reported later by Jacovella and coworkers \cite{jacovella2016a}. The state selected ions were not spatially separated and remained in the supersonic beam. 

Compared to electron bombardment and REMPI, the MATI method allows selective preparation of ionic states with arbitrary rovibronic excitation. Exotic states, inaccessible from the ground state, can be prepared using a multiphoton, stepwise resonant excitation scheme, by maximizing the overall Franck-Condon factors. The highest vibrational levels of the electronic ground state and the rovibrational levels of the first excited electronic state of H$_2^+$, HD$^+$ and D$_2^+$ can be prepared in such a way \cite{beyer2016b,beyer2018b}. These excited states can be beneficial in precision measurements, because they offer an enhanced sensitivity for fundamental constants, such as the proton-to-electron mass ratio \cite{augustovicova2014a}, or peculiar effects like ortho-para mixing \cite{moss1993c,critchley2001a,beyer2018c}. 

In this paper, we review the MATI mechanism and discuss additional requirements to efficiently extract only the state-selectively prepared ions from the supersonic beam and inject them into a radio-frequency ion trap, which offers long interaction times and good control of environmental conditions. Based on designs for EBIT sources, we analyze a system containing a dc quadrupole bender and ion lenses to deflects the ion packet out of the supersonic beam and focuses it into the trap \cite{zitzer2024a, moritz2025a, eierman2023a, schmoger2015a, asvany2014a, redshaw2023a, sels2020a, hilger2013a}.  The results presented in this paper are general to ions of any charge-to-mass ratio.

\section{Principle of mass-analyzed threshold ionization}
\begin{figure}[htbp!]
    \centering
    \includegraphics[width=1.0\linewidth]{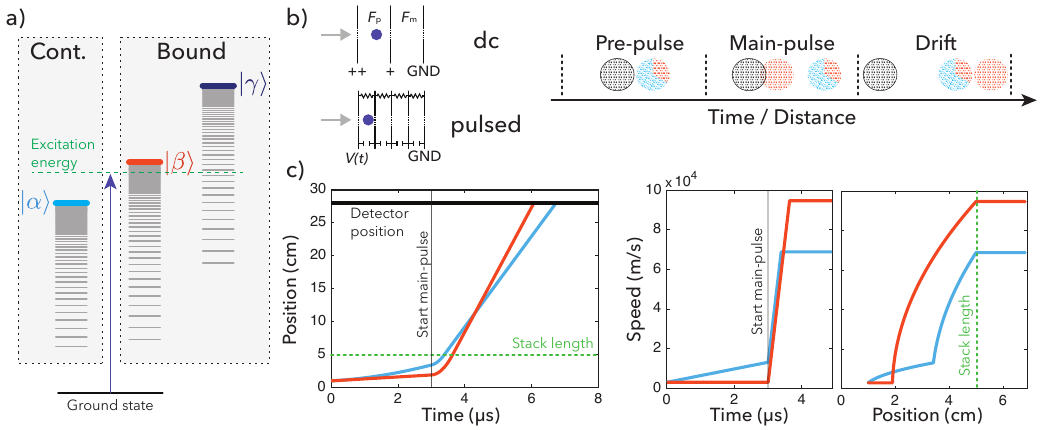}
    \caption{Principle of Mass-Analyzed Threshold Ionization (MATI). a) Energy-level scheme for the photoexciation with Rydberg series converging to three ionic levels $\ket{\alpha}, \ket{\beta}, \ket{\gamma}$. The laser is slightly red-detuned from the $\ket{\beta}$ ionization limit. b) Separation of prompt and MATI ions for dc and pulsed MATI. The grey arrow indicates the molecular beam and the dark blue circle the excitation region. The composition of the ion clouds is indicated by color, $\ket{\alpha}$ (dotted, blue) vs $\ket{\beta}$ (dotted, red), and the relative position of the neutral molecules (dotted, black) is shown. c) Trajectories for pulsed MATI of the prompt (blue) and MATI (red) ions. See text for details.}
    \label{fig:mati}
\end{figure}

Like ZEKE, MATI spectroscopy relies on the electric field ionization of long lived high-$n$ Rydberg states. An electric field $F$ (in V/cm) can lower the ionization threshold by $\approx 4.8 \sqrt{F}$~\wn \cite{hollenstein2001b}. \Figref{fig:mati}a shows the laser excitation from a ground or intermediate state into a region above the lowest ionization threshold with three ionization limits $\ket{\alpha}$, $\ket{\beta}$ and $\ket{\gamma}$, each representing the threshold of a Rydberg series. At the indicated excitation energy one obtains $\varepsilon\ell\ket{\alpha}$, $n_\beta\ell\ket{\beta}$ and $n_\gamma\ell\ket{\gamma}$, where $\varepsilon/n~\ell$ indicates a continuum or bound (Rydberg) state with orbital angular momentum $\ell$, respectively. With the laser red-detuned by about 2~\wn from threshold $\ket{\beta}$, $n_\beta \approx 250$ and $n_\beta \gg n_\gamma$.
The so-called \emph{prompt ions} in state $\ket{\alpha}$ are produced immediately through direct photoionization. The lifetime of the bound Rydberg states scales with $n^3$ and the states can decay by fluorescence, predissociation or autoionization. Only autoionization will produce molecular ions and because the Rydberg channel of $\ket{\beta}$ is energetically closed, these ions can only be formed in the state $\ket\alpha$. 
The high-$n$ Rydberg states $n_\beta\ell\ket{\beta}$ can be additionally stabilized through (i.) stray electric fields, for example through the prompt ions, which will lead to $\ell$-mixing and (ii.) collisions leading to $m$-mixing. Each of these effects adds another factor of $n$ to the lifetime, so that Rydberg states with $n>150$ typically have lifetimes in the microsecond range.

A small electric field, here called \emph{pre-pulse}, is applied first and it will sweep the prompt ions out of the excitation region and field-ionize the highest Rydberg states lying in the range of $\approx 1$~\wn~ below the $\ket{\beta}$ ionization threshold. This ion cloud, which contains particles in state $\ket{\alpha}$ and $\ket{\beta}$, respectively, is spatially separated from the (neutral) high-n Rydberg states. 
A second pulse, the \emph{main pulse}, field-ionizes the remaining $n_\beta\ell\ket{\beta}$ Rydberg states, selectively producing ions in the state $\ket{\beta}$. The interaction between the Rydberg electron and the molecular ion core is negligible in high-$n$ Rydberg states, so that the quantum state of the ion will be conserved during the pulsed-field ionization. The $\ell$-mixing further diminishes any remaining interaction with the ion core by adding character of the high-$\ell$ non-penetrating Rydberg states to the low-$\ell$ Rydberg states, typically accessible in field-free photoexciation. The absence of field-induced autoionization is apparent by the fact that ion signals are only observed in MATI within the narrow energy range of field ionization below each of the ionic thresholds \cite{mackenzie1995a}. At this point three particle packages exist, as indicated in \figref{fig:mati}b: the remaining neutral particles (black), the state-selected ion cloud, here called \emph{MATI ions} (red), and the prompt ion cloud (blue/red). The low-$n$ Rydberg states converging to the $\ket{\gamma}$ threshold are potentially subject to autoionization into the $\ket{\alpha}$ continuum, either under field-free or field-induced conditions (different parity and angular momentum selection rules apply). Under typical conditions, the ions produced in that way will be part of the prompt-ion cloud. The situation becomes more complicated when the ionization thresholds $\ket{\beta}$ and $\ket{\gamma}$ are within the MATI resolution of approximately 1~\wn, an example being the hyperfine structure of H$_2^+$ ($\sim0.03$~\wn). In this case, Stark-induced mixing will prevent any discrimination of ions between the two thresholds. While MATI based methods cannot be used to create molecular ions in selected hyperfine states, this might be possible by pulsed-field ionization of hyperfine-resolved Rydberg states \cite{beyer2018e}. 

In MATI spectroscopy, the pre- and main-pulse, as well as the time delay between them, will influence the total time-of-flight and need to be optimized so that the prompt and MATI ion signal do not overlap, as shown in \figref{fig:mati}c. From a practical point, it is advantageous to have the smaller MATI ion pulse arrive before the larger prompt ion pulse to avoid saturation of the detector. The drift region is essential for obtaining a large difference in the arrival time on the detector. In contrast, the maximum difference in the speed is obtained at the end of the stack, or when the main pulse is turned off.

Experimentally, both dc and pulsed implementations are possible. For dc MATI, a Wiley-McLaren type setup with three electrodes/meshes can be used \cite{johnson1994a}. In the simplest case, this defines two regions with electric fields $F_p$ and $F_m$, for the pre- and main-pulse, respectively. The velocity of the neutral molecules in the supersonic beam in combination with the lengths of the two regions sets the timescale. 
For pulsed MATI, a set of electrodes is combined using a voltage-divider network with impedance matching \cite{hollenstein2001b}, so that a homogeneous electric field can be applied over a large volume. After a field-free excitation, the pre-pulse $F_p$ is applied for a duration $t_p$, followed by the main-pulse $F_m$. 

A typical experimental setup for MATI spectroscopy employs pulsed dye lasers with nanosecond pulse lengths and can operate at repetition rates between 1 and 1000~Hz and produces between 1-100 ions per shot. These repetition rates for the ion production surpass the typical cycling rates for the ion trap precision measurements. It can be expected that the separation between prompt and MATI ions becomes more challenging with increasing mass, but MATI has been applied successfully even for heavier molecules. The mass and the ion velocity also control the timescale needed for the ion transport from the source region to the ion trap. For heteronuclear molecules this might mean that black-body-induced redistribution must be taken into account.

\subsection{Ionization Rates \& Survival Fractions}\label{sec:ionization_rates}
\begin{figure}[htbp!]
    \centering
    \includegraphics[width=1.0\linewidth]{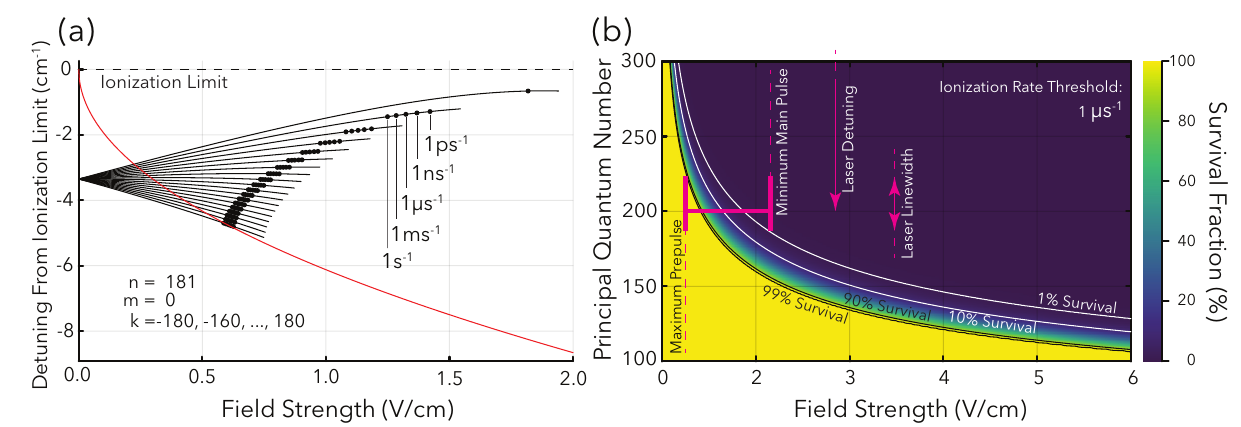}
    \caption{a) Stark map for the ($n=181$, $m=0$) manifold. Only every 20th Stark component is shown for clarity. Various field-ionization rates are indicated for the individual components by dots. The red line indicates the classical saddle-point ionization threshold. b) Survival fraction of Rydberg states as a function of the electric field. For all shown calculations, the Rydberg constant $R_\infty$ is used, without the reduced mass correction.}
    \label{fig:StarkFigure}
\end{figure}

The field-ionization rates of the high-$n$ Rydberg states must be known to evaluate the effect of the pre- and main-pulse. Maximally efficient state preparation with MATI ionizes as few Rydberg molecules as possible with the pre-pulse $F_p$, but maximizes the spatial separation and the velocity difference of the prompt and MATI ion clouds.

Following the procedure outlined in Ref.~\cite{hollenstein2001b}, we calculated the Stark-shifted energies and ionization rates for a given field strength $F$, based on a perturbative treatment for the hydrogen atom \cite{damburg1979a}. \Figref{fig:StarkFigure}a shows the simplified Stark map for the manifold with $n=181$ and $m=0$ (there is a total of 181 substates, but only every 20th component is plotted, centered at the $n_1-n_2=0$ component). For a single principal quantum number, high- and low-field seeking states are field ionized at significantly different field strengths. The classical saddle-point ionization threshold \cite{littman1978a}, predictive of the earliest-ionizing sublevels, is plotted in red. The ionization rate increases exponentially with $F$ \cite{damburg1979a}, resulting in a negligible probability that a given sublevel has an ionization rate comparable to the experimental timescale. This means that pulsed field-ionization can be assumed to be instantaneous on the timescale of the ion trajectories. We assume a threshold ionization rate of $10^6\si{\per\second}$, and molecules that have a calculated ionization rate greater than the threshold are assumed to be instantly field ionized.

\Figref{fig:StarkFigure}b shows the survival fraction of the Rydberg level population, characterized by the principal quantum number $n$, as a function of the electric field strength. For a fixed field, it can be seen that the survival fraction sharply increases from $<1\%$ to $>99\%$. Indicated by red arrows are the detuning ($\sim3$~\wn) and the line width ($\sim1$~\wn) of the photoexcitation laser. The maximum pre-pulse amplitude is determined by having a maximum $1\%$ loss of the highest Rydberg states through ionization by the pre-pulse and the minimum main-pulse field strength is given by requiring field-ionization of $99\%$ of the original population. The smallest difference between pre- and main pulse for a given range of Rydberg states (corresponding to a horizontal band in \Figref{fig:StarkFigure}b, on the order of $\sim 2$~V/cm) is caused by the significantly different field-ionization rates for low- and high-field seeking states. In principle, the main-pulse could be made arbitrarily strong, however, the kinetic energy of the ions gained during PFI must be removed for injecting them into a trap.

The transition probability for the Rydberg excitation scales as $n^{-3}$, and the level density scales as $n^3$, canceling each other out in the regime where the spacing between Rydberg levels is much smaller than the linewidth of the laser. This leads to a constant photoexcitation cross section for the quasi-continuum of the high-$n$ Rydberg states.

\section{Energy Ratio via Dimensionless Parameters}\label{sec:energy_contrast}
For traditional spectroscopic applications of MATI, the difference in the total flight time (in the MATI stack and the drift region) between the prompt and MATI ions is maximized \cite{johnson1994a}. We consider a MATI configuration, where the electric fields are applied parallel to the direction of the supersonic beam. As will be discussed in Section \ref{sec:injection} this proves to be advantageous for injecting the ions, given the small transverse velocity spread. Applying MATI to ion generation, the figure-of-merit is the energy ratio $\kappa\equiv v_\text{MATI}^2/v_\text{prompt}^2$, as a large energy ratio improves the performance of the ion extraction. The kinetic energy of the prompt ions is calculated assuming constant acceleration during the pre- and main pulses, as long as the ions remain within the separation region. For MATI ions, only the main pulse leads to acceleration. In general terms, the pre-pulse will set the relative positions of prompt and MATI ions at the onset of the main pulse, which in turn determines the total distance traveled under the acceleration of the main pulse field.  

To simplify our treatment, we assume:
\begin{itemize}
    \item The MATI takes place with a uniform electric field within the separation region, which extends from $-L_r$ to $L_f$ around the photoexcitation point.
    \item Ionization occurs instantly at field strengths that lead to an ionization rate of $> 10^6\si{\per\second}$. See subsection \ref{sec:ionization_rates} for details.
    \item The initial velocity and position spreads of the excited molecules are neglected.
    \item Space-charge effects are neglected.
\end{itemize}

Under these assumptions, a clear optimum in energy ratio is found in both accelerating and retarding pre-pulse cases.

\begin{figure}[htbp!]
    \centering
    \includegraphics[width=\linewidth]{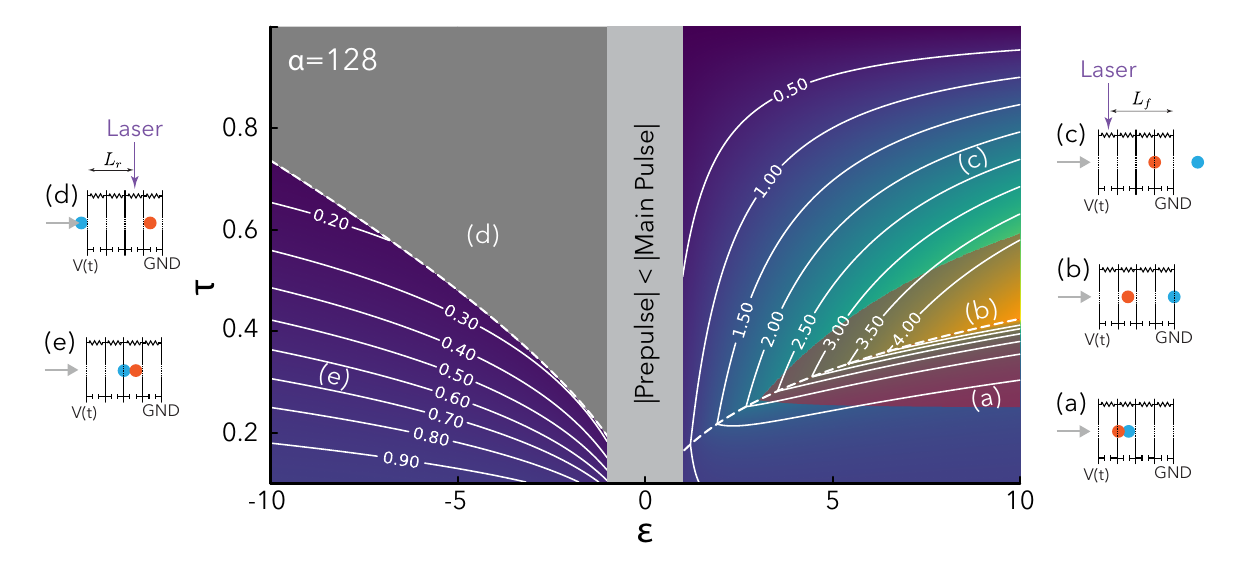}
    \caption{The energy ratio $\kappa$ (indicated by the contour lines) in the $\epsilon$-$\tau$ parameter space with a constant $\alpha=128$. For points (a)-(e), diagrams show the relative positions of the prompt (blue) and MATI (orange) ions at the moment the main pulse is applied. The dashed lines indicate when the prompt ions have left the separation region.  For the $\epsilon>0$ and $\epsilon<0$ cases, the laser is drawn in different locations so that the relative positions of the prompt and MATI ions can be exaggerated for the purposes of illustration. See text for details.}
    \label{fig:nondim_contours}
\end{figure}

The initial velocity of the neutral beam $v_0$, the duration of the pre-pulse $t_p$, the distance from the laser excitation to the edge of the uniform field region $L_{r,f}$, the mass of the molecule $m$, the charge of the resultant ion $q$, and the respective pre-pulse and main pulse field strengths $F_p$ and $F_m$ are expressed with three dimensionless parameters,
\begin{equation*}
    \tau\equiv\frac{t_p v_0}{L_f},\qquad 
    \epsilon\equiv\frac{F_m}{F_p}, \qquad
    \alpha\equiv\frac{qF_mL_f}{\frac{1}{2}mv_0^2}\ .
\end{equation*}

$\tau$ is the ratio of the drift distance of the neutral molecules during $t_p$ compared to the effective separation region length. MATI is only defined between $\tau\ge0$ (no pre-pulse) and $\tau\le1$ (Rydberg states leave the separation region and are not field ionized). $\epsilon$ represents the ratio of the main and pre- pulse field strengths and for MATI we impose $\left|\epsilon\right|>1$. The ratio of potential energy caused by the main pulse ($t_p$=0) to the kinetic energy of the neutral molecules is given by $\alpha$. $\alpha$ represents the upper limit of the kinetic energy that the MATI ions will gain and it is assumed that $\alpha>1$.  It represents the upper limit because any experimental MATI implementation involves a delay between the pre-pulse and the main pulse, during which time the state-selectively excited molecules will drift forward, reducing the total length over which they will be ultimately accelerated during the main pulse.  $\alpha$ is distinct from $\kappa$ in that $\kappa$ represents the actual energy ratio arising from a MATI sequence, whereas $\alpha$ corresponds to the hypothetical energy gain factor that would be achieved if the MATI ions were field-ionized and accelerated immediately after Rydberg excitation.

The dimensionless parameterization of the MATI sequence shows that the mass of the state-selectively excited molecule emerges in two different parameters, explicitly in the denominator of $\alpha$, but also implicitly in $\tau$ because of the $v_0$ dependence in the numerator. For a supersonic beam, one finds $v_0 \propto \sqrt{1/m}$ \cite{haberland1985a}, implying that heavier molecules drift more slowly. Optimal values of $\tau$ then require either a shorter separation region length $L_f$, or longer pre-pulses (this can increase the radiative loss). The parameter $\alpha$ becomes independent of the mass due to the cancellation in the denominator ($mv_0^2\propto m^0$). When a gas mixture is used, the dependence of $v_0$ on $m$ breaks down and the only generalization that can be made is the scaling of $\tau$.

\subsection{Accelerating Pre-Pulse, $\epsilon>1$}
The results are divided into two sub-cases depending on whether the prompt ions remain (regime a, $\tau<\tau^*$) or are ejected (regime c, $\tau\ge\tau^*$) from the separation region during the pre-pulse, as shown in \figref{fig:nondim_contours}. $\tau^*$ is obtained as $\tau^*=2\frac{-\epsilon+\sqrt{\epsilon(\epsilon+\alpha)}}{\alpha}$.  The optimal (largest) energy ratio always occurs at the boundary (dashed line, case b in \figref{fig:nondim_contours}) for $\tau=\tau^*$. This corresponds to the prompt ions exiting the separation region at the same moment that the main pulse turns on.

The energy ratio is given by
\begin{equation*}
    \kappa=
    \begin{cases}
    \frac{4 \epsilon ^2 (\alpha(\tau-1)-1)}{\alpha^2 \tau^2 (\epsilon -1)-4 (\alpha +1) \epsilon ^2+4 \alpha\tau \epsilon(\epsilon -1) } & \tau<\tau^* \vspace*{1em}\\
    \frac{\epsilon(1+\alpha-\alpha\tau)}{\alpha+\epsilon} & \tau\ge\tau^*
    \end{cases}\, .
\end{equation*}

\subsection{Retarding Pre-Pulse, $\epsilon < -1$}
For the retarding pre-pulse, two sub-cases can be defined depending on whether the prompt ions remain (regime e) or are ejected (regime d) from the separation region before the main pulse arrives. In the latter case, the energy ratio is formally not defined, but experimentally this corresponds to perfect separation of the prompt and MATI ions, both traveling in opposite directions. 
The prompt ions are able to explore the part of the separation region upstream of the laser excitation.  We introduce the length of this region using the parameter $r\equiv L_r/L_f$, as a ratio.
The dashed line in the left side of \figref{fig:nondim_contours} indicates the boundary, for which $\tau_r^*=2\frac{-\epsilon+\sqrt{\epsilon(\epsilon-r\alpha)}}{\alpha_r}$. 
For $\tau_r<\tau_r^*$, 
\begin{equation*}
    \kappa=\frac{4 \epsilon ^2 (\alpha(\tau-1)-1)}{\alpha^2 \tau^2 (\epsilon -1)-4 (\alpha +1) \epsilon ^2+4 \alpha\tau \epsilon(\epsilon -1) }\ .
\end{equation*}
The prompt ions have a larger kinetic energy than the MATI ions, $\kappa<1$, because they are accelerated by the main pulse over a longer distance. The energy ratio improves monotonically ($\kappa$ decreases) with increasing $\tau$. 

A retarding pre-pulse is more efficient for short pre-pulse durations (see \figref{fig:nondim_contours} for $\tau\lesssim0.3$). The parameter space where the accelerating pre-pulse is more efficient is shaded red (between regime a and b in figure \ref{fig:nondim_contours}).

\subsection{Reducing the Energy Spread and Enhancing the Energy Ratio}\label{sec:enhancement}
For molecules that have undergone Rydberg excitation, the initial position distribution, dictated by the spatial extent of the excitation laser, is the dominant cause of energy spread within the MATI ion ensemble. A different initial position in the separation region directly translates into a different kinetic energy gained by the ions during the main pulse. This can be improved by switching the main pulse off, while all MATI ions are still within the electrode region. In this case, the energy gain for all ions is set by the MATI pulse duration and not by the traversed distance. 

By decelerating the prompt and MATI ions in a later stage, the energy ratio between the prompt and MATI ions and within each group of ions can be enhanced.

\subsection{DC MATI}
The previous analysis for pulsed MATI can be applied for the dc case, by identifying that for $\tau=1$ the Rydberg molecules have reached the end of the first ($F_p$) segment. The kinetic energy gain, $\Delta K_2>0$, in the second ($F_m$) segment is the same for both prompt and MATI ions. In turn, the energy ratio at the end of the first segment, $\kappa=\epsilon/\left(\alpha+\epsilon\right)$, decreases monotonically with increasing $\Delta K_2$, according to $\kappa=(\Delta K_2+\frac{1}{2}mv_0^2)/(\Delta K_2+\frac{1}{2}mv_{prompt}^2)$. 
For a retarding pre-pulse, the preferred configuration corresponds to regime (d) in \figref{fig:nondim_contours}, where the prompt ions are ejected in the opposite direction, achieving a perfect separation of the prompt and MATI ions.
dc MATI might prove advantageous for continuous molecular beams.

\section{Energy-Selective Elements and Ion Injection}
Implementation of an energy selection scheme preferably takes advantage of ion optics elements that are already used in experiments that inject externally produced ions.
DC Quadrupole benders \cite{zeman1977a} and Einzel lenses are almost universally found in existing experiments that involve ion injection into linear Paul traps, Penning traps, and also occasionally for injection into reflectron mass spectrometers \cite{zitzer2024a, moritz2025a, eierman2023a, schmoger2015a, asvany2014a, redshaw2023a, sels2020a, hilger2013a}.
The former in particular is promising for serving a dual purpose, separating the prompt and MATI ions from the neutral beam but also isolating the MATI ions based on their specific kinetic energy.
We find that the energy selectivity of these existing beamline elements is sufficient, eliminating the need for dramatic changes in existing experiments to accommodate specialized ion beam monochromators.

\subsection{Quadrupole Bender}
The quadrupole bender \cite{zeman1977a} is a ion optics element that bends a beam by $90^\circ$. The bend takes place in the plane of an electrostatic quadrupole field, with the beam entering along the tangent to the equipotential surface midway between the hyperbolic electrodes. Chromatic aberration, the energy dependence of the optical properties, serves to separate the prompt and MATI ions by exploiting their energy ratio.

\begin{figure}[htbp!]
    \centering
    \includegraphics[width=1\linewidth]{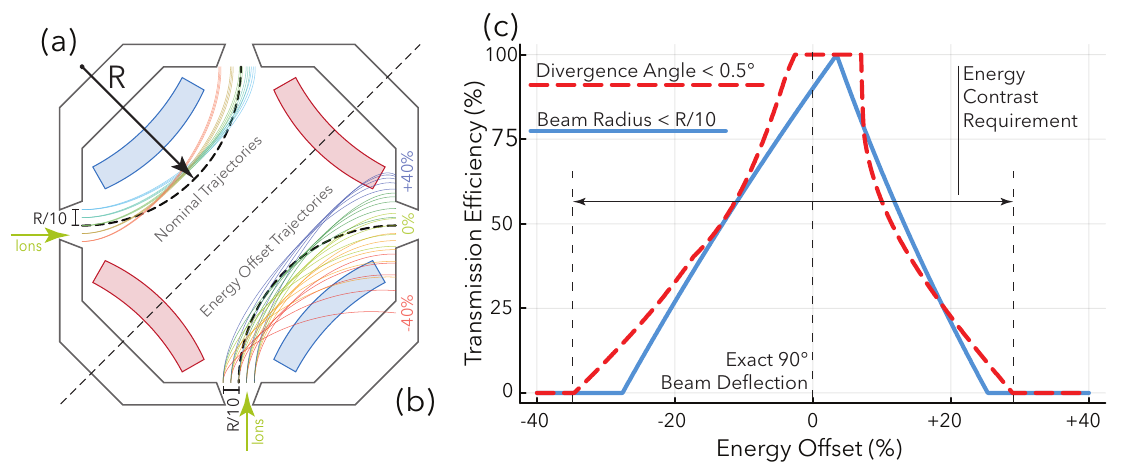}
    \caption{Nominal trajectories at $0\%$ energy offset are shown in (a), where a ray that enters along the incident beam's axis leaves the bender exactly at the center.  Each color denotes a specific offset from the incident beam's axis.  In (b), trajectories for incident particles at differing energies (denoted by the colors) are shown, with each color corresponding to the location of an axis tick in (c), which shows typical energy-selectivity characteristics of a quadrupole bender.  Depending on the criterion used to define the acceptance of the quadrupole bender, broadly similar characteristics are obtained.}
    \label{fig:bender_acceptance}
\end{figure}

\Figref{fig:bender_acceptance} shows the general scheme for the use of an idealized quadrupole bender as a monochromator, to separate the MATI ions from the prompt ions.
As seen in \figref{fig:bender_acceptance}(a-b), the deflection of the positively-charged ion beam arises as a combination of repulsion from the positively-charged electrodes and attraction to the negatively-charged electrodes. The quadrupole bender is, to first order, insensitive to the ion energy (first-order achromatic) and offset from the nominal trajectory. The energy selectivity arises only as a second-order effect, giving a tradeoff between selectivity and beam quality. A first-order energy selector would be able to distinguish between prompt and MATI ions, that have very close kinetic energies, but the MATI ion beam would be degraded by the energy spread within the MATI ion cloud.

With a first-order achromatic energy selector, the relatively small energy spread of the MATI ion beam causes negligible distortion, but the separation in kinetic energies between the prompt and the MATI ions must also be larger.
\figref{fig:bender_acceptance}(c), shows that a perfect separation of the two beams requires an energy ratio of about $30\%$, which, see \figref{fig:nondim_contours}, can be easily achieved for $\kappa<0.7$ or $\kappa>1.3$.
The limited energy selectivity of the quadrupole bender is compensated by the large energy ratio that can be achieved by optimizing the MATI operating point.

The selectivity of the quadrupole bender is only dependent on the kinetic energy per unit charge, not on the mass of the ion. This is seen by reparameterizing the equation of motion using $\tau\equiv t/\sqrt{m}$
\begin{equation*}
    m\frac{d^2\, \vec{r}}{dt^2}=\frac{d^2\, \vec{r}}{d\tau^2}=-q\, \vec{\nabla}\cdot\Phi(\vec{r})\ .
\end{equation*}
Furthermore, the cation's charge $q$ enters only as a multiplicative factor to the potential, which must anyways be tuned so that the principal ray emerges at a right angle.  Therefore, the results that are presented for the quadrupole bender are general for any mass or (nonzero) charge.

To simulate the energy-dependent acceptance of the quadrupole bender without complicating the analysis with too many device-specific effects, such as fringing fields, an ideal quadrupole potential was assumed.
The size of the incoming beam was set as $1/10$ of the bending radius, $R$, and for the purposes of producing sub-figures (a) and (b), the incoming beam was considered to be perfectly collimated.
The equations of motion for the ions are numerically solved, and the final state of the ion, once it has reached the boundary of the square simulation domain, is checked to determine if it satisfies the acceptance condition.
We chose two sample acceptance conditions: (i.) for the solid blue line, the transmission efficiency is defined as the proportion of the incident ions that remain within an offset of $R/10$ from the beam axis and (ii.) the dashed red line is obtained by finding the fraction of particles that have an angle of less than $0.5^\circ$ from the beam axis.

\subsection{Einzel Lenses}
In an Einzel lens, separation is not achieved via chromatic aberration but by setting the central electrode potential so that prompt ions lack the energy to pass through. However, using an Einzel lens alone in the beamline is unfavorable, as it keeps state-selected ions in the neutral beam path, raising the risk of ion–neutral collisions, and links the lens’s focusing properties to the required energy resolution.

\subsection{Injection}\label{sec:injection}
For axial injection of ions into a linear Paul trap it is important to focus the ion beam to a small area at the trap entrance, minimizing the effect of the rf fields. The initial radial displacement will define the amplitude of the transverse motion in the ion trap. An ion beam that can be tightly focused should have a small beam radius and a small transverse velocity spread (i.e., being collimated). Both of these requirements are satisfied by MATI, because the phase-space volume of the state-selected ions (the beam emittance) is relatively small.
A skimmed supersonic beam of neutral molecules inherently has a tight transverse and longitudinal velocity distribution. The spatial extent of the ion cloud is given by the intersection of the molecular and laser beams, selecting an ensemble of excited molecules that is narrowly distributed simultaneously in space and velocity.

\begin{figure}[htbp!]
    \centering
    \includegraphics[width=\linewidth]{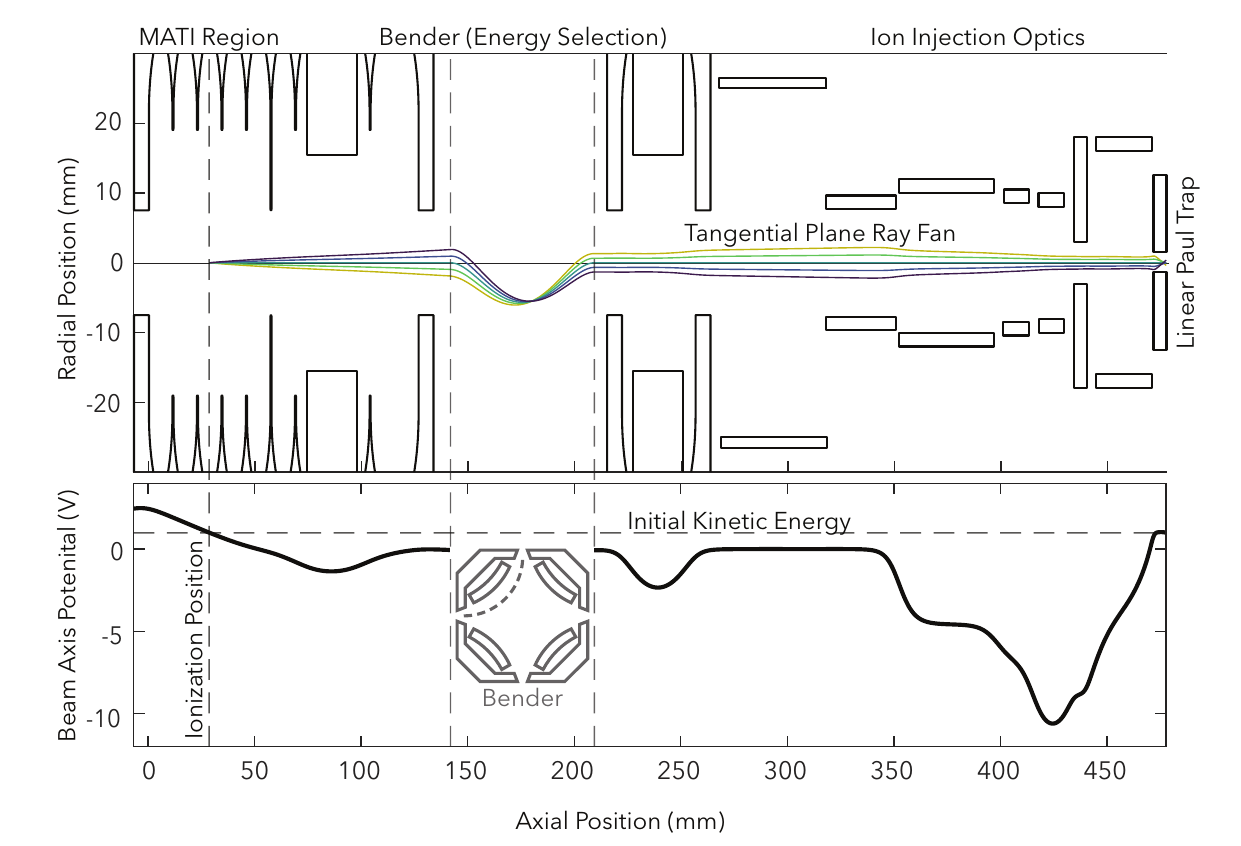}
    \caption{MATI based ion injection system. Upper panel: electrode cross-sections and ray fan plot showing ion trajectories in the tangential plane of the apparatus. The radial position of the ions is always taken with respect to the beam axis, which is taken to be circular within the quadrupole bender to smoothly transition between the two perpendicular sections of the apparatus. Lower panel: Potential energy of the MATI ions along the instrument axis.}
    \label{fig:injection}
\end{figure}

A proposed scheme for state-selective ion preparation using MATI, energy selection with a quadrupole bender, and injection of the resultant ions into a linear Paul trap is shown in the upper panel of \figref{fig:injection} \cite{knapp2026b}. The system consists of an electrode stack for pulsed MATI, a quadrupole lens before and after the quadrupole bender, and an objective lens assembly. The lower panel of \figref{fig:injection} indicates the potential energy that the ions will experience while traversing the system. The initial kinetic energy (indicated by a dashed line) that the ions will gain during the MATI sequence sets the dc bias of the ion trap. The value is chosen to be slightly lower to guarantee a low velocity of the ions when entering the trap.

\begin{figure}[htbp!]
    \centering
    \includegraphics[width=\linewidth]{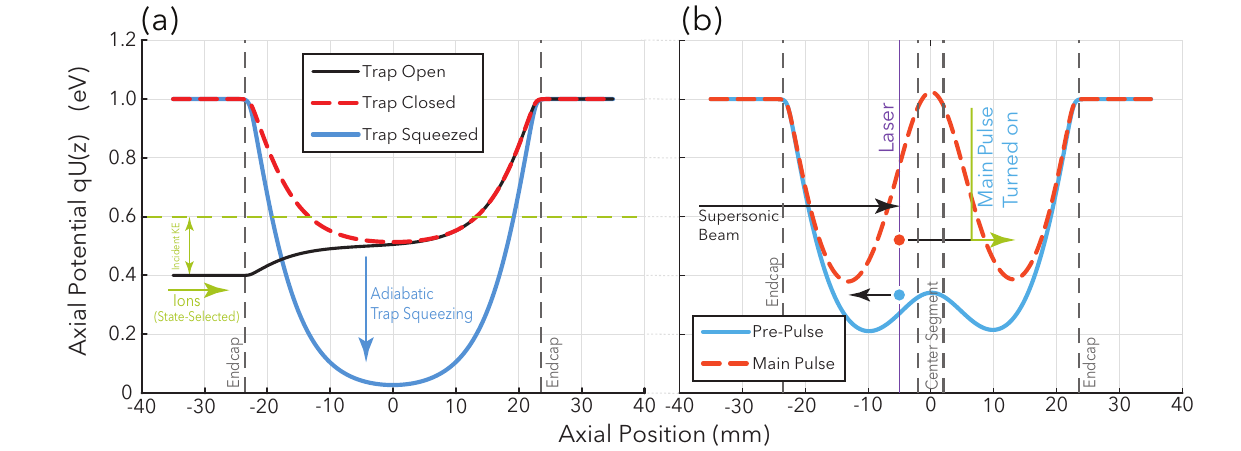}
    \caption{The potentials associated with (a) injection of externally state-selected ions into an ion trap and (b) a scheme to accomplish MATI directly within an ion trap.}
    \label{fig:trap_potentials}
\end{figure}

Admitting the ions is accomplished by temporarily lowering the entrance endcap potential, as depicted in \figref{fig:trap_potentials}(a). During the injection, a shallower axial trap depth is preferred so that the ions take longer to rebound from the far endcap electrode. Switching the endcap on with the ions in the vicinity would accelerate them.
If a stronger confinement is required, the relative dc bias between the rf quadrupole electrodes and the endcaps should be slowly increased.

\subsection{MATI Inside of an Ion Trap}
Compared to the dedicated MATI setup as described before, in a linear Paul trap, a strong rf electric field is present. Under typical operating conditions, this field is sufficient to field-ionize the high-$n$ Rydberg states \cite{zhang2023a}.  
MATI within an ion trap is challenging because of the necessity to distinguish between the prompt ions and the state-selected ions, rather than indiscriminately trapping everything that was ionized within the trap.
Nonetheless, within segmented traps that allow modulation of the axial potential, the endcap potentials may be used to perform MATI using a supersonic beam that is directed axially into the trap.
\Figref{fig:trap_potentials}(b) shows an example of such a scheme.
The laser is aligned to be slightly in front of the central segment, such that the prompt ions are pushed backward upon being ionized, and are trapped in the first segment of the trap.
The Rydberg molecules are allowed to drift further towards the second segment, and the field ionization is achieved by suddenly increasing the potential applied to the central dc electrode.
This ionizes the molecules, trapping them within the second trap segment.

Written in terms of the Mathieu $q$ and $a$ parameters \cite{paul1990a} for an rf angular frequency $\Omega$, the field amplitude as a function of $r$ is given by
\begin{equation}\label{eq:Fmax_dirty}
    \left| F_{\text{max}}\right|=\frac{m\Omega^2}{2Ze}\left(q+\frac{\left|a\right|}{2}\right)r\ .
\end{equation}
Here, $q$ (the Mathieu parameter) is taken to be positive, $m$ represents the mass of the ion species being trapped, $Z$ its charge state, and $e$, the elementary charge. This parametrization removes the dependence on the trap radius.
The period of the rf drive is much faster than the time that a neutral molecule takes to drift into another segment, so the above formula is an accurate representation of the maximum electric field that would be seen by a radially displaced molecule in a Rydberg state.

In the $a=0$ case (no static quadrupole field component), this expression can be further simplified to eliminate other apparatus- and ion-specific constants.  In the secular ($q\ll1$) limit, the effective trap depth $U_\text{eff}$ in volts (defined as the time-averaged potential energy of a test charge at a radial displacement equal to the trap radius $R$) is given as
\begin{equation*}
    U_\text{eff}\approx\frac{q^2}{16}\frac{m\Omega^2R^2}{Ze}\ .
\end{equation*}
The maximum field strength at radius $r$, per unit trap depth is obtained:
\begin{equation*}\label{eq:Fmax_clean}
    \frac{\left| F_{\text{max}}\right|}{U_\text{eff}}=\frac{8}{q}\frac{r}{R^2}\ .
\end{equation*}
This expression loses its dependence on ion-specific parameters such as $m$ and $Z$ because the ion trap drive voltage amplitude and frequency share the same dependence on them when we impose that the trap depth is the same.

In the case of an H$_2^+$ in a $10$~MHz linear Paul trap, operating at $q=0.1$ and $a=0$, the resultant field gradient is roughly $40.9$~(V/cm)/ mm, using equation (\ref{eq:Fmax_dirty}).  The trap depth given by these parameters with a trap radius $R=5$~mm is approximately $1.28$~V.  For heavier ions, the trap parameters can be varied (the rf amplitude and/or the rf frequency reduced) to keep the same $1.28$~V trap depth, resulting overall in the same $40.9$~(V/cm)/ mm field gradient.  This result is therefore general to ions of any charge-to-mass ratio.

The steep field gradient imposes the requirement that the excitation volume (intersection of molecular and laser beams) be a fraction of a millimeter in radius, or alternatively, that the trap must be operated at an extremely weak confinement during the MATI sequence by reducing the Mathieu parameter $q$ (or $\Omega$, if at all experimentally feasible).
MATI within an ion trap is therefore not entirely impossible, but it poses formidable experimental challenges, especially for heavier ions that require larger rf amplitudes to confine. 

\section{Conclusion}
By parameterizing the MATI sequence with a pre- and main pulse, an optimal energy ratio between MATI and prompt ions can be identified for a range of experimental conditions. We have shown how standard ion-optical elements can then be used to isolate the MATI ions and outlined the requirements for efficient axial injection into a linear rf trap. An experimental implementation for the state-selective preparation of H$_2^+$ ions is currently under construction, and the accompanying code enables adaptation of these results to other use cases.

\section{Code Availability Statement}
The code for producing the results shown in figures \ref{fig:StarkFigure}, \ref{fig:nondim_contours}, and \ref{fig:bender_acceptance} is provided in the form of a \verb|Pluto| notebook, implemented in the programming language \verb|Julia|.  It can be found at the following URL/DOI: \url{https://doi.org/10.5281/zenodo.18670846}

\section*{Acknowledgments}
It is a particular pleasure to dedicate this article to Frédéric Merkt on the occasion of his 60th birthday. He introduced one of us (M.B.) to molecular spectroscopy, Rydberg physics and, most importantly, to molecular hydrogen. His exemplary scientific rigor, intellectual curiosity and generosity as a mentor have been a continual source of inspiration. We are deeply indebted to him for shaping our scientific outlook and for many stimulating discussions over the years. It is an honor to contribute this work in recognition of his outstanding contributions to science and to our community. 

\section*{Disclosure Statement}
No potential conflict of interest was reported by the author(s).

\section*{Funding}
This research was supported by the Dutch Research Council (NWO) through a VENI (VI.Veni.202.140) and an ENW-M2 (OCENW.M.21.141) grant.

\bibliographystyle{tfo}

\section{Appendices}

\noindent\textbf{Appendix A. Design Procedure}\medskip
\subsubsection{General Scheme}

A design procedure to optimize the operating point of a MATI system is presented below, assuming $L_f=L_r$ for simplicity.
If perfect separation of the ions is possible by making $L_r$ sufficiently small, the analysis below is unnecessary.
\begin{enumerate}  
    \item Choose a laser detuning and linewidth. These parameters create an upper bound on the pre-pulse field strength ( \figref{fig:StarkFigure}(b)) and a lower bound on the main pulse field strength.
    Key considerations include
    \begin{enumerate}
        \item efficient use of the laser's spectral density so that not too much of the power goes to photons with energies higher than the ionization threshold, and
        \item avoidance of interlopers that may reduce the lifetimes of nearby Rydberg levels.
    \end{enumerate}
    If possible, a larger red-detuning from the ionization threshold is preferred because this gives more flexibility in choosing a larger pre-pulse field strength $F_p$ (radiative decay may become prohibitive).
    \item Choose main pulse field strength $F_m$.
    Several possible considerations may set an upper bound for $F_m$, including
    \begin{enumerate}
        \item the intended final kinetic energy of the state-selectively prepared MATI ions, and
        \item field-induced autoionization.
    \end{enumerate}
    Within these limitations, $F_m$ should be chosen to be as large as possible, to obtain better energy ratio.

    \item Choose the pre-pulse field strength $F_p$.
    A full analysis requires a numerical calculation of the ionization rate for each excited sublevel, but the classical 'saddle-point' threshold can be used as a rough heuristic.
    The highest Rydberg level prepared by the excitation laser should not be field-ionized by the pre-pulse.
    The pre-pulse should always be chosen with the largest possible magnitude, as this gives a monotonic improvement for the energy ratio.
    The only question that remains is whether to choose a retarding pre-pulse $\epsilon<0$, or an accelerating pre-pulse $\epsilon>0$.
    This choice should be made in the final step.
    At this point, the magnitude but not the sign of $\epsilon$ is determined.
    
    \item Choose a separation region length $L$ to obtain $\alpha$.  This may be an iterative process that involves carrying out the rest of the design procedure to determine if the chosen value is suitable.
    A longer $L$ (corresponding to an increasing value of $\alpha$) results in a larger energy ratio $\kappa$ for a given $\epsilon$, because the state-selected ions are accelerated over a longer distance. However, this comes with the tradeoff that a longer pre-pulse duration $t_p$ is required to bring the prompt ions to the end of the region before applying the main pulse.
    
    \item  The final step involves the determination of $\tau$, and the determination of the sign of $\epsilon$ (that is, whether a retarding or an accelerating pre-pulse is chosen).
    For the accelerating pre-pulse case, choose the final operating point to be
    \begin{equation}\label{eq:optimum_tau}
        \tau=\min\left(\tau_\text{max},\ \tau^*\right)\ ,
    \end{equation}
    where $\tau_\text{max}$ corresponds to the longest experimentally viable pre-pulse duration, and the other option ($\tau^*$) is the global optimum for energy ratio, corresponding to the condition that the prompt ions exit the stack just as the main pulse is applied.
    $\tau_{\text{max}}$ is practically set by the lifetime of the state-selectively excited Rydberg levels.
    For the retarding pre-pulse case, $\tau=\tau_\text{max}$ is the optimal operating point because the energy ratio improves monotonically.
    However, no additional improvement is achieved by using a longer prepulse, beyond the point $\tau>\tau_r^*$, where the ions are ejected from the back of the separation region.
    The $\pm\epsilon$ cases should be compared to choose the final operating point.
    For reference, the parameter space where $\epsilon>0$ gives a superior energy ratio is highlighted in figure \ref{fig:nondim_contours}.
\end{enumerate}

\subsubsection{Example Application 1: H$_2^+$}
As an example, we consider the state-selective production of rovibrational levels of H$_2^+$ (laser linewidth of $0.1$~\wn).
\begin{enumerate}
    \item We choose a laser detuning of about $-2.7$~\wn from the ionization threshold, corresponding to excitation around $n=200$, where a large S/N-ratio is expected \cite{beyer2018e}.
    Rydberg levels in the principal quantum number range $n=193,\dots,207$ are included within two linewidths from the center, and we consider this to be the manifold of state-selectively excited Rydberg levels (this is a slightly smaller range than in the exaggerated illustration for figure \ref{fig:StarkFigure}(b)).
    
    \item The main pulse field strength $F_m$ must be sufficient to ionize all of the excited Rydberg levels.
    The $n=193$ manifold sets the lower limit, and inspection of figure \ref{fig:StarkFigure} (a) shows that a field of about $F_m\ge2.0~\si{\volt\per\centi\meter}$ should be used to efficiently ionize all of the levels.
    Although a larger field would give better results in practice, we choose $F_m=2.0~\si{\volt\per\centi\meter}$ for the purposes of determining the worst-case scenario.
    
    \item The $n=207$ manifold of Rydberg sublevels ionizes at the lowest field strength, so these effectively set the upper bound for the pre-pulse field strength, giving roughly $\left|F_p\right|\le0.34~\si{\volt\per\centi\meter}$.
    At this point, we have $\left|\epsilon\right|=2.0~\si{\volt\per\centi\meter}/0.34~\si{\volt\per\centi\meter}\approx 5.9$
    
    \item We assume an upper bound of $10~\si{\micro\second}$ for the pre-pulse duration that would realistically be used.
    The supersonic molecular beam of neutral H$_2$ has an initial speed of roughly $3000~\si{\meter\per\second}$, so the desired pre-pulse duration implies that the separation region length $L$ should be about $3~\si{\centi\meter}$ long at minimum, but the stack should be made longer so that the MATI ions can still be accelerated after the full pre-pulse duration has been used.
    Therefore, we choose $L=6~\si{\centi\meter}$.
    Note that H$_2$, being the lightest molecule, is somewhat of an extreme case in this respect, and heavier molecules will require significantly shorter separation regions.
    Overall, we have $\alpha=12~\si{\electronvolt}/0.094~\si{\electronvolt}\approx128$.

    \item We have already chosen $\tau_\text{max}=0.5$ by doubling the drift distance of the neutral beam at the maximum pre-pulse duration to obtain $L$.
    With this choice, substituting into equation (\ref{eq:optimum_tau}), we find the optimum operating point for the pre-pulse duration to be $\tau=0.35$, the global maximum for the chosen $\alpha$ and $\left|\epsilon\right|$, corresponding to a $7~\si{\micro\second}$ pre-pulse for the chosen operating parameters.
    This gives an energy ratio factor of $\kappa=3.7$, comfortably more than even the loose bounds for a quadrupole bender as a kinetic-energy-selective element (figure \ref{fig:bender_acceptance}).
    Smaller $\tau$ are also acceptable, for instance, for $\tau=0.25$, an energy ratio of approximately $\kappa=1.5$.
    At $\tau=0.25$, the accelerating pre-pulse is only marginally better than the retarding pre-pulse, assuming $L_f=L_r$.
    
\end{enumerate}

\subsubsection{Example Application 2: O$_2^+$}
Steps (1) through (3) are identical to the above H$_2^+$ case if the quantum defect is neglected.
\begin{enumerate}
  \setcounter{enumi}{3}
  \item The supersonic beam of O$_2$ is expected to have an initial speed of roughly $750~\si{\meter\per\second}$.  Taking the same $10~\si{\micro\second}$ upper bound on the pre-pulse duration, we obtain $L=7.5~\si{\milli\meter}$ long at minimum, but including the allowance for additional acceleration after the full pre-pulse duration, we take $L=1.5~\si{\centi\meter}$.  Overall, we have $\alpha=(2~\si{\volt\per\centi\meter})(3~\si{\centi\meter})e/0.094~\si{\electronvolt}\approx32$.  (The resultant kinetic energy of the molecules in the supersonic beam is the same.)
  \item We obtain $\tau^*=0.57$, but for the design we stipulated that $\tau_{\text{max}}=0.5$.  This indicates that the separation region is too short to obtain the optimum energy ratio.  We can move forward and use the full $10~\mu$s allotment for the pre-pulse duration, or by lengthening to $L=2.5~\si{\centi\meter}$, $\tau^*=0.48$ (corresponding to $9.6~\mu$s) can be achieved because the lengthened region gives $\alpha=53$.
\end{enumerate}
If $10~\mu$s is found to be too long of a duration for the prepulse, a long separation region can be used to enhance $\alpha$.  For instance, taking $L=5~\si{\milli\meter}$, we find $\tau^*=0.74$ and the optimal pre-pulse duration is reduced to $5~\mu$s.  The energy ratio suffers, with $\kappa=1.33$ in this configuration, at the lower limit of the quadrupole bender's energy resolution (at least with the acceptance conditions we use as examples in the main text).

It is worth investigating the retarding prepulse case.  Take $L_r\to0$ to to obtain the most advantageous configuration for ejecting the prompt ions backwards (the distance to the back is minimized to ensure ejection with the shortest possible prepulse).  Then $\tau_r^*=-4\epsilon/\alpha$, and since the state-selectively excited molecules must stay within the separation region, $\tau_r^*<1\implies \alpha>-4\epsilon=23.6$.  This is achieved with $L_f>1.1~\si{\centi\meter}$.  However, this represents the extreme case where the MATI ions are field ionized at the exact moment that they leave the separation region, at the exact moment where the prompt ions leave the separation region at the opposite side, a situation that is not experimentally viable.  Taking a small margin and setting $L_f=1.5~\si{\centi\meter}$, we find that $\tau_r^*=0.74$, corresponding to a pre-pulse duration of $15~\mu$s, much longer than what is achievable using the accelerating pre-pulse.  We conclude that the accelerating pre-pulse is strongly preferred.

\end{document}